\documentclass[12pt]{article}
\usepackage{graphicx,color,epsfig,rotate}

\title{Ferromagnetism in the Hubbard model with long-range 
and correlated hopping}
\author{Pavol Farka\v sovsk\'y\\
Institute  of  Experimental  Physics,  Slovak   Academy   of
Sciences\\
Watsonova 47, 043 53 Ko\v {s}ice, Slovakia}
\date{}
\begin{document}
\baselineskip=24pt
\maketitle

\begin{abstract}

The extrapolation of small-cluster exact-diagonalization
calculations is used to examine ferromagnetism in the one-dimensional 
Hubbard model with long-range and correlated hopping. It is found that the 
correlated hopping term stabilizes the ferromagnetic state for a wide 
range of electron interactions $U$ and electron concentrations $n$. 
The critical value of the interaction strength $U_c$ above which 
the ferromagnetic state becomes stable is calculated numerically and 
the ground-state phase diagram of the model is presented 
for physically the most interesting cases.

\end{abstract}

\newpage
The Hubbard model has become, since its
inception~\cite{Hubbard} in 1963, one of the most popular examples
of a system of interacting electrons with short-range interactions.
It has been used in the literature to study a great variety of many-body
effects in metals, of which ferromagnetism, metal-insulator transitions,
charge-density waves and superconductivity are the most common
examples. Of all these cooperative phenomena the problem of ferromagnetism
in the Hubbard model has the longest history. Although the model was
originally introduced to describe the band ferromagnetism of transition
metals, it soon turned out that the single-band Hubbard model is not
the canonical model for ferromagnetism. In fact the existence of saturated
ferromagnetism has been proven rigorously only for very special
limits. The first well-known example is the Nagaoka ferromagnetism
that comes from the Hubbard model in the limit of infinite repulsion
and one hole in a half-filled band~\cite{Nagaoka}.
Another example where saturated ferromagnetism has been shown to
exist is the case of the one-dimensional Hubbard model with
nearest and next-nearest-neighbor hopping at low electron
densities~\cite{M_H}.
Moreover, several examples of the fully polarized ground state have been
found on special lattices (special conduction bands)
as are the fcc-type lattices~\cite{Ulmke},
the bipartite lattices with sublattices
containing a different number of sites~\cite{Lieb},
the lattices with unconstrained hopping of electrons~\cite{Pieri}
and the flat bands~\cite{M_T}. This indicates that the lattice structure 
and the kinetic energy of electrons, i.e., the type of hopping play 
an important role in stabilizing the ferromagnetic state.  

In our previous paper~\cite{Fark1} we have shown that if the electron hopping 
is described by a more realistic model (than the nearest-neighbor hopping)
then ferromagnetism comes from the Hubbard model naturally
for a wide range of the model parameters. No extra interactions
terms should be included. In particular, we have found that the
long-range hopping with power decaying hopping amplitudes $t_{ij}$
given by~\cite{Fark2}

\begin{equation}
t_{ij}(q)=\left \{ \begin{array}{ll}
  \quad 0,               &   i=j,\\
  -q^{|i-j|}/q,          &   i\neq j
\end{array}
\right.
\end{equation}
gives rise to ferromagnetism for electron densities above half-filling.
As soon as $q$ that controls the effective range of the hopping
($0\le q \le 1$) is different from zero the ferromagnetic state 
is stabilized for all Coulomb interactions $U$ greater than 
some critical interaction strength $U_c(q)$ that value is 
dramatically reduced with increasing $q$. From this point of view 
one of main reasons why the ferromagnetic state absent in  the ordinary
Hubbard model with nearest-neighbor hopping ($q=0$) is that the description 
of the electron hopping was too simplified.  

Here we further generalize this model by introducing the correlated
hopping term, in which the $\sigma$-electron hopping amplitudes between 
lattice sites $i$ and $j$ depend explicitly on $n_{i-\sigma}$ 
and $n_{j-\sigma}$ occupancy, i.e. 

\begin{equation}
t^{\sigma}_{ij}=t_{ij}[1+t'(n_{i-\sigma}+n_{j-\sigma})].
\end{equation}

The importance of the correlated hopping term on the ground-state 
properties of the Hubbard model has been already mentioned by 
Hubbard~\cite{Hubbard}. Later Hirsch~\cite{Hirsch} pointed
out that this term may be relevant in explanation of 
superconducting properties of strongly correlated electrons.
Here we examine effects of this term on the stability of 
the fully polarized ferromagnetic state. The same subject has been 
studied recently by Amadon and Hirch~\cite{Amadon}, as well as by
Kollar and Vollhardt~\cite{Kollar}, however 
they considered hopping only between the nearest-neighbor sites.

The selection of hopping matrix elements in the form given by 
Eq.~1  and Eq.~2 has several advantages. It represents a much more 
realistic type of electron hopping on a lattice (in comparison to
nearest-neighbor hopping), and it allows us to change
continuously the type of hopping and thus immediately study the effect 
of long-range and correlated hopping on the ground state properties 
of the Hubbard model.

The Hamiltonian of the single-band Hubbard model in which the effects 
of long-range and correlated hopping are incorporated is given by

\begin{equation}
H=\sum_{ij\sigma}t^{\sigma}_{ij}c^+_{i\sigma}c_{j\sigma}+
U\sum_{i}n_{i\uparrow}n_{i\downarrow},
\end{equation}
where $c^+_{i\sigma}$ and $c_{i\sigma}$ are the creation and annihilation
operators  for an electron of spin $\sigma$ at site $i$,
$n_{i\sigma}$ is the corresponding number operator
($N_{\sigma}=\sum_i n_{i\sigma}$) and
$U$ is the on-site Coulomb interaction constant.

To examine possibilities for existence of ferromagnetism in this model
the ground states are determined by exact diagonalizations for a 
wide range of model parameters ($q,t',U,N=\sum_{\sigma}N_{\sigma} $). 
Typical examples are then chosen from a large 
number of available results to represent the most interesting cases.
The results obtained are presented in the form of phase diagrams 
in the $U$-$q$ plane. To determine the phase diagram in the $U$-$q$
plane (corresponding to some $t'$ and $N$) the ground state energy 
of the model is calculated point by point as functions of $q$ and $U$. 
Of course, such a procedure demands in practice a considerable amount 
of CPU time, which imposes severe restrictions on the size of 
clusters ($L$) that can be studied with this method~($L \sim 16$).
Fortunately, we have found that the ground-state energy of the 
model depends on $L$ only very weakly (for a  wide  range of the 
model parameters) and thus already such small clusters can describe 
satisfactorily the ground state properties of the model.

The results of our small-cluster exact-diagonalization calculations
obtained on finite clusters up to $L=16$ sites are summarized in 
Fig.~1 and Fig.~2. There is shown the critical interaction strength $U_c$, 
above which the ground state is ferromagnetic, as a function of $q$ for 
selected values of $n=N/L$ and $t'$ ($n=3/2,7/4; t'=0,0.2,0.4$). 
To reveal the finite-size effects on the stability of ferromagnetic domains, 
the behavior of the critical interaction strength $U_c(q)$ has been calculated
on several finite clusters at each electron filling. It is seen that 
finite-size effects on $U_c$ are small and thus these results can be 
satisfactorily extrapolated to the thermodynamic limit $L\to \infty$.           
Our results clearly demonstrate that the ferromagnetic state is strongly 
influenced by correlated hopping ($t'$) and generally it is stabilized with 
increasing $t'$. The effect is especially strong for intermediate and strong 
values of $q$. Even, there exists some critical value of $q$ above which 
the ground state 
is ferromagnetic for all nonzero $U$. With increasing $t'$ this critical
value shifts to lower values of $q$ (that represent a much more realistic 
type of electron hopping) and the ferromagnetic domain correspondingly
increases. Performing exhaustive numerical studies of the model for a wide
range of electron concentrations (on different lattice clusters) we have
found that the model exhibits the same behavior for all electron
concentrations above half-filling~\cite{note}, and that with increasing
concentration this effect becomes more pronounced (see Fig.~1 and Fig.~2).
These results clearly show that ferromagnetism comes naturally from the 
Hubbard model with long-range and correlated hopping for a wide range of model 
parameters without any other assumptions.  This opens a new  route towards 
the understanding of ferromagnetism in the Hubbard model.

In summary, the extrapolation of small-cluster exact-diagonalization
calculations was used to examine ferromagnetism in the one-dimensional 
Hubbard model with long-range and correlated hopping. It was found that the 
correlated hopping term stabilizes the ferromagnetic state for a wide 
range of electron interactions $U$ and electron concentrations $n$. 
The critical value of the interaction strength $U_c$ above which 
the ferromagnetic state becomes stable was calculated numerically and 
the ground-state phase diagram of the model is presented 
for physically the most interesting cases.

\vspace{0.5cm}
This work was supported by the Slovak Grant Agency VEGA
under Grant No. 2/7021/20 and the Science and Technology 
Assistance Agency under Grant APVT-51-021602. Numerical 
results were obtained using computational resources of 
the Computing Centre of the Slovak Academy of Sciences.

\newpage
Figure Captions

\vspace{0.5cm}
Fig. 1. The critical interaction strength $U_{c}$  as a function 
of $q$ calculated for different $t'$ and $L$ at $n=3/2$. 
Curves from up to down correspond to: $t'=0, 0.2$ and 0.4.

\vspace{0.5cm}
Fig. 2. The critical interaction strength $U_{c}$  as a function 
of $q$ calculated for different $t'$ and $L$ at $n=7/4$. 
Curves from up to down correspond to: $t'=0, 0.2$ and 0.4.

\newpage
\begin{figure}[htb]
\hspace{-2cm}
\includegraphics[angle=0,width=18.0cm,scale=1]{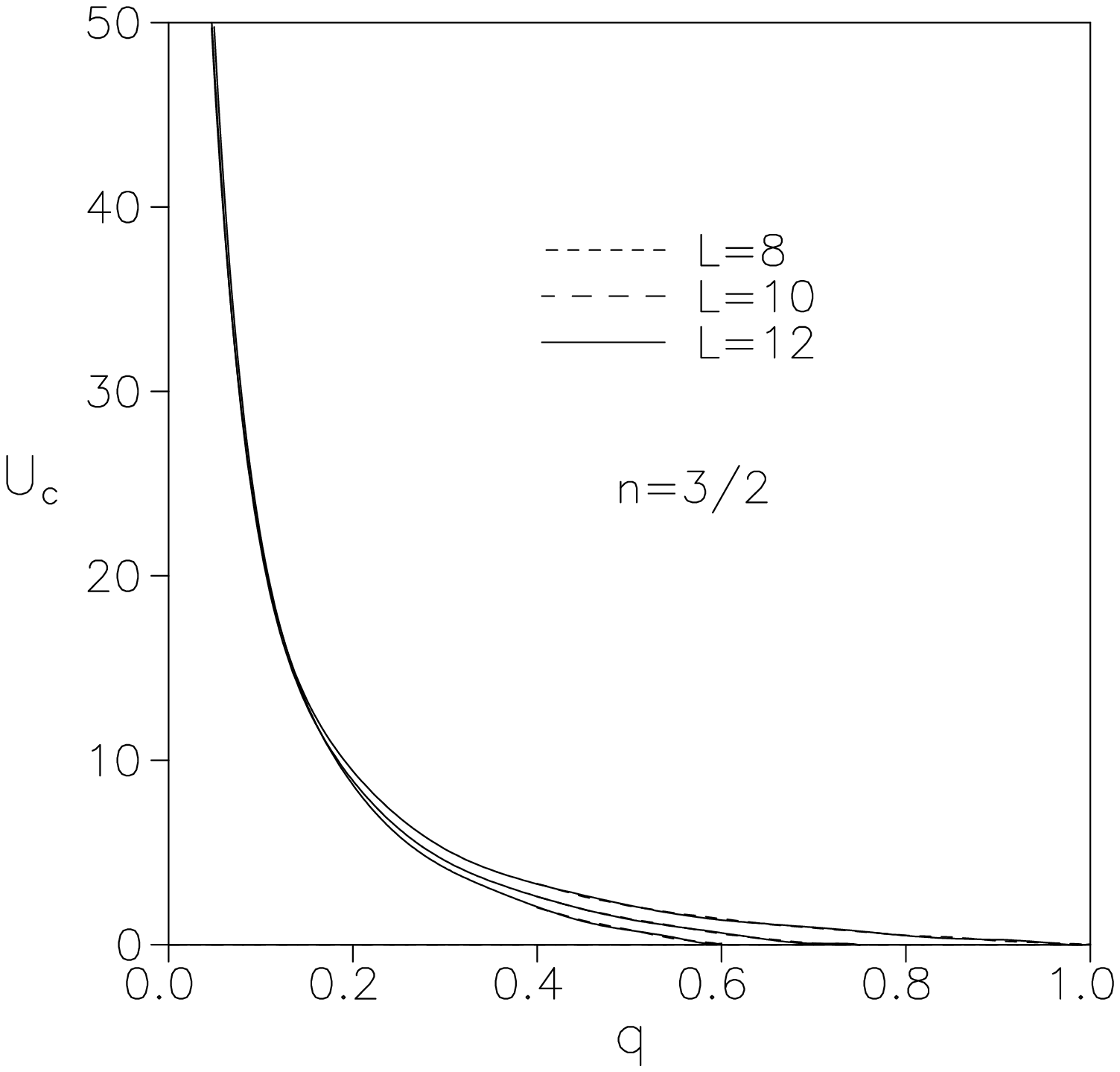}
\caption{ }
\label{fig1}
\end{figure}

\newpage
\begin{figure}[htb]
\hspace{-2cm}
\includegraphics[angle=0,width=18.0cm,scale=1]{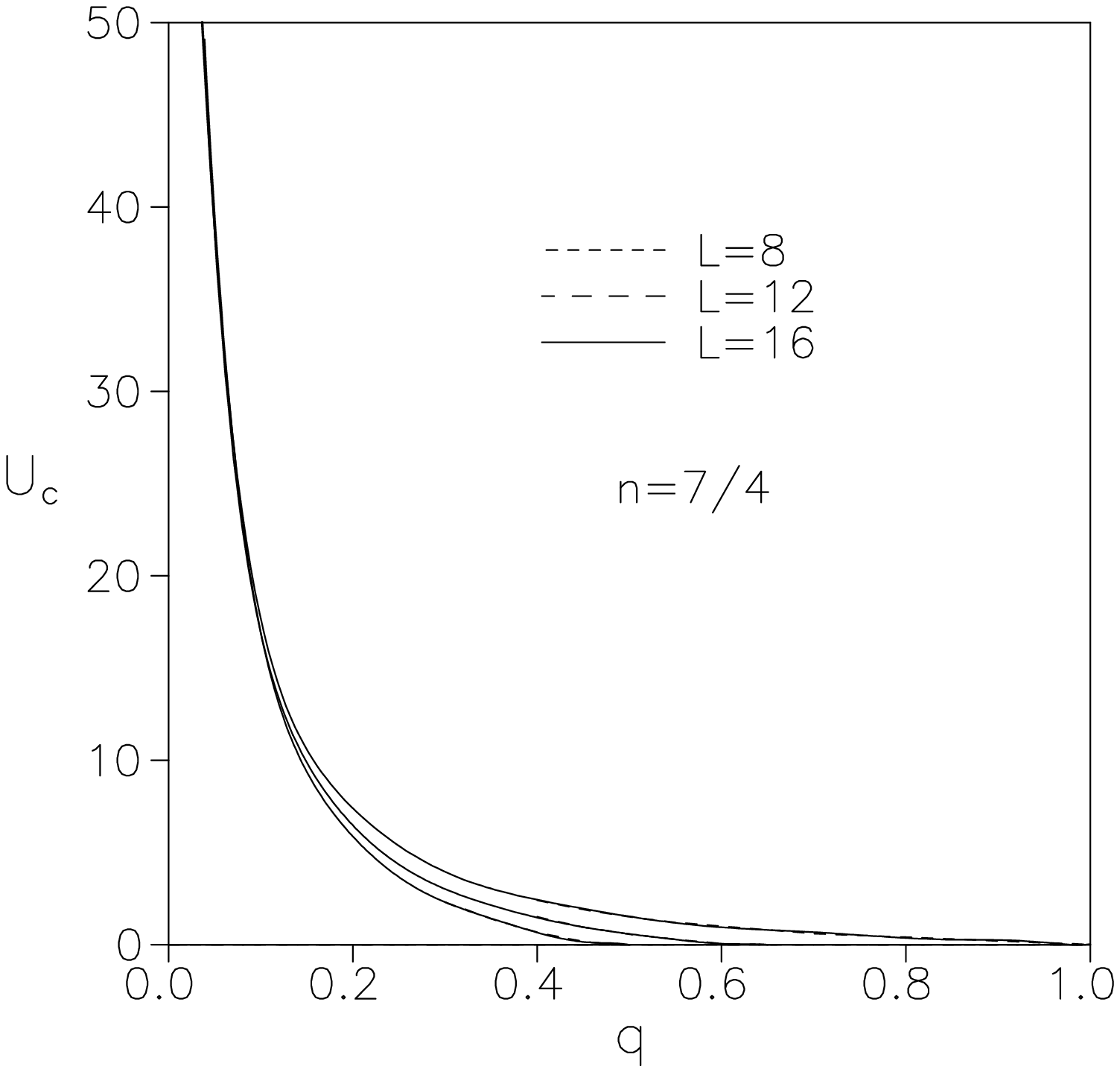}
\caption{ }
\label{fig2}
\end{figure}


\newpage
\begin{thebibliography}{99}

\bibitem{Hubbard} J. Hubbard, Proc. R. Soc. London A {\bf 276}, 238 (1963).

\bibitem{Nagaoka} Y. Nagaoka, Phys. Rev. {\bf 147}, 392 (1966).

\bibitem{M_H} E. M\'uller-Hartmann, J. Low. Temp. Phys.{\bf 99}, 342 (1995).

\bibitem{Ulmke} M. Ulmke, Eur. Phys. J. B {\bf 1}, 301 (1998).

\bibitem{Lieb} E.H. Lieb, Phys. Rev. Lett.  {\bf 62}, 1201 (1989).

\bibitem{Pieri} P. Pieri,
Mod. Phys. Lett. B{\bf 10}, 1277 (1996); 
M. Salerno, Z. Phys. B {\bf 99}, 469 (1996);
{\it ibid}. B {\bf 101}, 619 (1996).

\bibitem{M_T} A. Mielke and H. Tasaki, Commun. Math. Phys.  {\bf 158},
341 (1993).

\bibitem{Fark1} P. Farka\v{s}ovsk\'y, Phys. Rev. B {\bf 66}, 012404 (2002).

\bibitem{Fark2} P. Farka\v{s}ovsk\'y, J. Phys. Condens. Matter
{\bf 7}, 3001 (1995); {\it ibid.} {\bf 7}, 9775 (1995);
Phys. Rev. B {\bf 57}, 14722 (1998);
Int. J. Mod. Phys. B  {\bf 12}, 803 (1998).

\bibitem{Hirsch} J.E. Hirsch,  Physica C {\bf 158}, 236 (1989).

\bibitem{Amadon} J.C. Amadon and J.E. Hirsch,  Phys. Rev. B {\bf 54}, 6364
(1996).

\bibitem{Kollar} M. Kollar and D. Vollhardt,  Phys. Rev. B {\bf 63}, 045107
(2001).

\bibitem{note} $n\leq 1$ and $t'<0$ does not stabilize the ferromagnetic 
state.

\end{thebibliography}
\end{document}